# MAPPING COMPLIANCE: A TAXONOMY FOR POLITICAL CONTENT ANALYSIS UNDER THE EU'S DIGITAL ELECTORAL FRAMEWORK

Marie-Therese Sekwenz / Rita Gsenger

Marie-Therese Sekwenz PhD Candidate, TU Delft, Faculty of Technology Policy and Management, Building 31
Jaffalaan 5, 2628 BX, Delft , Netherlands
M.T.Sekwenz@tudelft.nl; https://www.tudelft.nl/en/staff/m.t.sekwenz/

Rita Gsenger Research Associate,Weizenbaum Institute, Research Group Norm-Setting and Decision Processes,
Hardenbergstr. 32, 10623 Berlin, DE; PhD Candidate, Faculty for Communication Sciences, Free University Berlin, Garyst. 55,
14195 Berlin
rita.gsenger@weizenbaum-institut.de; https://www.weizenbaum-institut.de/portrait/p/rita-gsenger/

*Keywords:*   ***Elections, Political campaigns, Advertisements, Digital Services Act***

*Abstract:*   *The rise of digital platforms has transformed political campaigning, introducing complex regulatory challenges. This paper presents a comprehensive taxonomy for analyzing political content in the EU's digital electoral landscape, aligning with the requirements set forth in new regulations, such as the Digital Services Act (DSA). Using a legal doctrinal methodology, we construct a detailed codebook that enables systematic content analysis across user-generated and political ad content to assess compliance with regulatory mandates.*

## 1. Introduction

After the 2016 US elections and the Brexit referendum, political consensus formed in the EU that novel rules for the context of elections are needed to increase online protection.[1] The digital dimensions of political debate and their potential negative consequences can have manifestations offline, which was illustrated by the storm on the US Capitol in 2021 or the attempt to replicate such behavior in Germany.[2] Governmental processes aim to protect voters, politicians, and political debates, including the obligation to decide how much regulation is needed for disinformation, digital campaigning, or content moderation. The European Union's key regulations and soft law provisions that regulate the digital side of elections include:

- The Digital Services Act (DSA),[3]
- The Regulation on Transparency and Targeting of Political Advertising (TTPA),[4]

---

[1] HALL/TINATI/JENNINGS, From Brexit to Trump: Social Media's Role in Democracy, Computer 51, p. 18; HENDRICKS/SCHILL, The Social Media Election of 2016, in Denton Jr (Hrsg.), The 2016 US Presidential Campaign: Political Communication and Practice, Springer, Cham 2017, p. 121.

[2] NG/CRUICKSHANK/CARLEY, Cross-platform information spread during the January 6th capitol riots, Social Network Analysis and Mining 12, p. 133; ASSOCIATED PRESS, „Anti-corona" extremists try to storm German parliament, The Guardian, https://www.theguardian.com/world/2020/aug/29/berlin-braces-for-anti-coronavirus-protest-against-covid-19-restrictions (accessed on 30.11.2024), 2020.

[3] Regulation (EU) 2022/2065 of the European Parliament and of the Council of 19 October 2022 on a Single Market For Digital Services and amending Directive 2000/31/EC (Digital Services Act) (Text with EEA relevance) OJ L 277.

[4] Regulation (EU) 2024/900 of the European Parliament and of the Council of 13 March 2024 on the transparency and targeting of political advertising (Text with EEA relevance).

- The Commission's Guidelines on Mitigation of Systemic Risks for Electoral Processes under the DSA (G-E–DSA).[5]

This article introduces key provisions relevant to the political context of online advertising and provides a codebook for content analysis in line with the DSA, TTPA, and G-E–DSA, aiming to support platforms' systemic risk assessments and external independent audits under Art. 34 DSA.

These regulations target different aspects and domains of political speech. The DSA mainly sets rules for content moderation and platform compliance with human rights and laws, demands regular risk assessments and audits, mitigation measures for risks, and creates obligations for some platforms to build ad repositories.

The TTPA includes norms on how campaigns and actors must be verified, which information must be included about spending and targeting, and ad repositories for platforms and for a new aggregated EU repository, according to Art. 13 TTPA. Therefore, understanding the interplay of these new regulatory measures is important for understanding the compliance obligations for platforms for user-generated or AI-generated political ad content. A recent example of the enforcement of novel regulations can be the opening of the proceedings against the Very Large Online Platforms (VLOPs according to Art 33 DSA) X (formerly known as Twitter) and TikTok.[6] X is investigated for potentially violating several provisions of the DSA, such as the risk on civic discourse and electoral processes (Art. 34(1) (c) DSA) or Art. 39 DSA laying down rules for online advertising transparency. Such violations can have severe consequences and could lead to fines of significant parts of the platforms' annual revenue, according to Art 74 DSA.

Therefore, platforms, advertisers, political candidates, campaigners, regulators, and civil society have to follow these regulations, which complement each other and provide additional details for specific use cases like generative AI.

In this article, we aim to introduce the key provisions relevant for the political context of online advertising to contribute to the ongoing debate about compliance with these regulations. Moreover, we provide a codebook for content analysis in line with the DSA, the TTPA, and the G-E–DSA. The codebook can be used to empirically test representative content samples of user-generated or ad-content to analyse systemic risks of Art 34 (1) DSA (mainly focusing on negative consequences on the electoral process according to Art 34 (1) (c) DSA).

Systemic risks that need to be assessed under the DSA can occur in various domains relevant to elections. For instance, risks can be situated between legal and harmful speech, mis- or disinformation, malicious flagging, Artificial Intelligence (AI) and deepfakes, targeting, or campaigns against candidates or specific groups. Disinformation is defined in the Code of Practice of Disinformation (CDP) as "verifiably false or misleading information [that is] created, presented and disseminated for economic gain or to intentionally deceive the public [and] may cause public harm [intended as] threats to democratic political and policymaking processes as well as public goods" and is an example of (potentially) harmful speech that not necessarily crosses the threshold of illegality.[7] Following the fine line between illegality and harmful speech, however, is posing questions to the protection of fundamental rights, such as freedom of speech or the right to free and fair elections.[8] Furthermore, if concepts like mis- or disinformation are not tangible and clearly definable,[9] the automatic detection of such violations can be hard to operationalise for automatic content moderation techniques like classifiers, or the use of Large Language Models (LLMs).[10]

---

[5] EUROPEAN COMMISSION, Commission on Guidelines for providers of Very Large Online Platforms and Very Large Online Search Engines on the mitigation of systemic risks for electoral processes pursuant to the Digital Services Act. https://digital-strategy.ec.europa.eu/en/library/guidelines-providers-vlops-and-vloses-mitigation-systemic-risks-electoral-processes (accessed on 19 November 2024), 2024.

[6] EUROPEAN COMMISSION, Commission opens formal proceedings against X under the DSA https://ec.europa.eu/commission/presscorner/detail/en/ip_23_6709 (accessed on 19 November 2024)), 2023; EUROPEAN COMMISSION, Commission opens formal proceedings against TikTok https://ec.europa.eu/commission/presscorner/detail/en/ip_24_926 (accessed on 19 November 2024), 2024.

To address the complex interplay of regulations and palette of risks to ensure free and fair elections of the digital age we will bring together three regulatory sources governing the election in the European Union (EU). By combining the three sources, we aim to provide a detailed overview of the demands for politicians, campaigners, advertisers, platforms, regulators, Trusted Flaggers, public authorities, interdisciplinary researchers (e.g., computer or social scientists), journalists, and other overseeing bodies in the EU.

In this text, we build on prior work for testing the systemic risk on the electoral process in the context of the German Federal elections in 2021.[11] For this article, we adapt our code book defining systemic risks under Art 34(1) (c) DSA for the notions needed to test the TTP, and the most recent G–E–DSA. To our knowledge, the selection of these legal sources adequately portrays the relevant regulatory landscape for the digital aspects of the electoral discourse and digital campaigning. This article aims to answer the following research questions:

- RQ1: What taxonomy can measure compliance with DSA, TTPA, and G–E–DSA to empirically assess systemic risks on the electoral process in political content?

- RQ2: What content annotation categories are necessary to be included in the content analysis process in line with the DSA, the TTP, and the G–E–DSA?

By answering these research questions, we will provide a consistent overview of applicable regulations in the context of elections in the digital realm and a method to comply with these regulations. This collection will contribute to understanding the complex interplay of the different regulatory sources and demands, helping advertisers, political campaigners, platforms, users, election overseeing bodies, regulators, and researchers better understand compliance at the intersection of the three regulatory sources and contribute with methodological tools to the scientific debate dealing with compliance measurement and automation of test and audits.

## 2. Political Advertising and Regulation

Political campaigns and the creation of political ads play an important role in elections. Politicians create comparatively cheap advertising strategies (in contrast to TV advertising) and fine-tune their message, visual, or appearance to target a specific voter audience, e.g., through gender filtering.[12] This separation between the political messages provided to the voters for different voter segments can have the consequence that not all voters see the same message of a political party.[13] This fragmentation, however, can create diverging political realities, e.g., through the distribution of generative-targeted AI content.[14]

---

[7] EUROPEAN COMMISSION, Code of Practice on Disinformation, 26.09.2018, (1).
[8] Charter of Fundamental Rights of the European Union OJ C 326.
[9] KAPANTAI ET AL, A systematic literature review on disinformation: Toward a unified taxonomical framework, New Media & Society 23, 5, 2021, p. 1301.
[10] BUCHANAN ET AL, Truth, Lies, and Automation: How Language Models Could Change Disinformation; VYKOPAL ET AL, Disinformation Capabilities of Large Language Models, ArXiv abs/2311.08838.
[11] KÜBLER ET AL, The 2021 German Federal Election on Social Media: Analysing Electoral Risks Created by Twitter and Facebook. Proc. of the 56th Hawaii International Conference on System Sciences (HICSS-56), Maui 2023.
[12] HOLMAN/SCHNEIDER/PONDEL, Gender Targeting in Political Advertisements, Political Research Quarterly 68, p. 816.

Risks to the electoral process might stem from the message conveyed by the ad content and the ad delivery process that decides who the target audience should be. In the ad delivery process, users might already be selected as a target audience who should be nudged to change their political opinion or form a voting decision.[15] Targeting in the ad display phase is also linked to the financial side of advertising. The price of ads usually includes an auctioning system defining different prices for users. For instance, users with different political leanings or affiliations can influence the price of the message presented in the ad (higher prices for presenting ads that do not match interest categories). Additionally, presenting the ad in different time frames can be relevant to prices (e.g., presenting the ad closer to the election date might have consequences on the pricing of the ad), or the targeting for different user groups can be a significant price indicator (of young mothers[16]). Nevertheless, financial data show that political players spend more than ever on digital campaigns.[17] How exactly such auctions work is not a transparent process and is only supported by aggregated numbers in repositories or transparency reports according to Art. 15, 24 and 42 DSA and Art. 39 DSA.

Attempts of new European regulations to make financial aspects of advertising more transparent include the indication of information in the user interface about high-level targeting criteria (for example, through the "why am I seeing this ad function") or through the mandatory ad repositories in the aggregated form of spending.[18]

By creating publicly accessible political ad repositories, more information about the electoral campaigns, the messages that politicians frame for specific audiences, visuals used, clips created, sponsors endorsed, or overall spending should create a better overview of the digital side of the election landscape. How meaningful transparency mechanisms are, however, is also very much related to the design of the tools used [36] and can be shifted to create different, more desirable visibilities.[19] Creating ad repositories for political content is now mandatory under the DSA and the TTPA on a platform level and for VLOPs and Very Large Search Engines (VLOSEs) on a European level (Art. 13 TTPA, Art 39 DSA). However, the completeness and accuracy of such repositories are criticized.[20] On the other hand, such inaccuracies could pose a potential systemic risk for platforms and their systemic risk assessments, risk mitigation obligations, and transparency reporting, according to Art. 35 DSA, Art. 39 DSA, Art. 15, Art 24, and Art 42 DSA.

---

[13] BELGIU/CONSTANTIN, Political Marketing Campaign, Logos Universality Mentality Education Novelty: Political Sciences & European Studies 4, p. 25.
[14] JINGNAN, AI-generated images have become a new form of propaganda this election season, NPR, https://www.opb.org/article/2024/10/18/ai-generated-propaganda-election-hurricane/ (accessed on 11 November 2024), 2024.
[15] ALI ET AL, Ad Delivery Algorithms: The Hidden Arbiters of Political Messaging, WSDM '21: Proceedings of the 14th ACM International Conference on Web Search and Data Mining, 2021.
[16] PACHILAKIS ET AL, YourAdvalue: Measuring Advertising Price Dynamics without Bankrupting User Privacy, Proceedings of the 2022 ACM SIGMETRICS/IFIP PERFORMANCE Joint International Conference on Measurement and Modeling of Computer Systems 2022, p. 41.
[17] SHEINGATE/SCHARF/DELAHANTY, Digital Advertising in U.S. Federal Elections, 2004-2020, Journal of Quantitative Description: Digital Media 2, 2022.
[18] FACEBOOK HELP CENTER, Why am I seeing ads from an advertiser on Facebook? https://www.facebook.com/help/794535777607370/ (accessed on 31 December 2024), 2024.
[19] FLYVERBOM, Transparency: Mediation and the management of visibilities, IJoC 10, 2016, p. 110.
[20] WHO TARGETS ME, A Goldilocks Zone for political ads https://whotargets.me/en/a-goldilocks-zone-for-political-ads/ (accessed on 18 November 2024), 2020.

However, it is very complex to define 'political content' on platforms (for example, what constitutes political advertising in cases of influencer content[21] and how to label or filter it automatically. Several platforms, such as TikTok, have claimed not to allow political content and failed.[22]

How can such claims be meaningfully and empirically tested in the light of new regulations? In empirical tests, false positives and negatives can shed light on the question of what constitutes good content moderation practices for identifying political ads. False positives can be illustrated as cases in which a piece of content is misidentified as political advertising. For instance, the ad of a building company is classified as political content. False negatives, therefore, are cases that do not classify political content as political. For example, a political party post advertising an event is not labeled as a political ad. The automation of these processes is an intersection with legal and contractual requirements like the Terms and Conditions (TaC) that has to be modeled and implemented by computer scientists to automate the processes needed for regulatory compliance.

Even as 'political content and advertising' on a VLOP like TikTok are prohibited in their TaC,[23] it is still relevant to ask how many false positives and false negatives are acceptable under new regulations and how to measure them.[24]

Besides, new risks may emerge for safeguarding political campaigns and voters online due to the increased use of generative AI. These risks are in the scope of the EU's AI Act.[25] Such risks could stem from the use of image generation in political contexts.[26] In the context of the campaigns for the US presidential election 2024, generative AI was used to generate the voice of the democratic candidate Joe Biden, to convince voters to 'save' their vote for the election in November and not vote in the primary elections.[27] On the other hand, deepfakes are another manifestation of the negative consequences of the digital sphere, or voice-generated content.[28] Moreover, these often have a gendered component. For example, Kübler et al. found that female politicians are targeted more often than male colleagues in disinformation-promoting content.[29]

Furthermore, different content forms of disinformation, such as deepfakes of audio content, can add complexity to the regulation and enforcement of the created rules; for example, how will the advertising repository for an audio- based platform be searchable and explorable in a meaningful way to spot risks early on?

---

[21] WHO TARGETS ME, How to correctly identify political ads (while acknowledging you can't). https://whotargets.me/en/how-to-correctly-identify-political-ads/ (accessed on 15 November, 2024) 2021.

[22] OREMUS, TikTok fails a test of its ban on political ads, Washington Post. https://www.washingtonpost.com/politics/2024/10/17/tiktok-political-ads-ban-report-global-witness/ (accessed on 12 November 2024), 2024.

[23] EUROPEAN COMMISSION, Commission launches new database to track digital services terms and conditions. https://digital-strategy.ec.europa.eu/en/news/commission-launches-new-database-track-digital-services-terms-and-conditions (accessed on 14 November, 2024), 2023.

[24] Delegated Regulation supplementing Regulation (EU) 2022/2065 of the European Parliament and of the Council, by laying down rules on the performance of audits for very large online platforms and very large online search engines, 20.10.2023.

[25] EUROPEAN COMMISSION, Commission on Guidelines for providers of Very Large Online Platforms and Very Large Online Search Engines on the mitigation of systemic risks for electoral processes pursuant to the Digital Services Act, 2024.

[26] VIGDOR, Trump Promotes A.I. Images to Falsely Suggest Taylor Swift Endorsed Him, The New York Times. https://www.nytimes.com/2024/08/19/us/politics/trump-taylor-swift-ai-images.html (accessed on 12 September 2024), 2024.

[27] ASSOCIATED PRESS, Company that sent fake Biden robocalls in New Hampshire agrees to $1m fine, The Guardian. https://www.theguardian.com/technology/article/2024/aug/22/fake-biden-robocalls-fine-lingo-telecom (accessed on 12 September 2024), 2024.

[28] MEAKER, Slovakia's Election Deepfakes Show AI Is a Danger to Democracy. Wired. https://www.wired.com/story/slovakias-election-deepfakes-show-ai-is-a-danger-to-democracy/ (accessed on 02 January 2025), 2023.

[29] KÜBLER ET AL, The 2021 German Federal Election on Social Media: Analysing Electoral Risks Created by Twitter and Facebook. Proc. of the 56th Hawaii International Conference on System Sciences (HICSS-56), Maui 2023.

However, not only digital risks may influence the political parquet before elections. For example, the platform Cameo was used as a vehicle in a disinformation campaign in the context of the ongoing Russian attack on Ukraine, instrumentalising American actors such as Elijah Wood to create individualised videos to create the impression of President Volodymyr Zelensky having a substance abuse problem.[30]

## 3. Introducing the legal provisions relevant for the context of elections and campaigning in the DSA, TTPA, and G–E–DSA

### 3.1. Ad Transparency and Systemic Risk in Europe's Digital Services Act

Under the DSA, VLOPs and VLOSEs are defined in Art. 33(1). They must test periodically the systemic risks defined in Art. 34 DSA and mitigate the defined risks according to Art. 35 DSA. Those platforms are measured against the threshold of reaching 45 million average monthly users within the EU. Their analysis of platform risks will be recorded in the audit report in line with Art. 42(4) (a) DSA and Rec. 100. These claims made in the audit report on the other hand will be tested by external independent auditors according to Art. 37 DSA. The mandatory risk assessments should cover five areas of audit according to Art. 34 (2) (a-e) DSA: the design of the algorithmic systems in use (including recommendation algorithms), the content moderation systems, the terms and conditions (including enforcement of the rules), systems for advertisement (including the creation of the ad, the audience selection tools and ad-delivery techniques), and data practices of VLOPs and VLOSEs. Advertising systems create a particular risk if run by VLOPs and VLOSEs according to Recital 95 DSA "for example in relation to illegal advertisements or manipulative techniques and disinformation with a real and foreseeable negative impact on public health, public security, civil discourse, political participation and equality."

In Art. 34, four risk categories are defined, laying down the testing frame for the internal systemic risk assessments, which have to be "identified, analyzed and assessed". To the context of elections, Art. 34(1) (c) DSA is most relevant for the context of elections, defining a systemic risk as "any actual or foreseeable negative effects on civic discourse and electoral processes, and public security." Additionally, to the problem domain of political content in the context of elections the exercise of fundamental rights as stated in Art 34 (1) (b) DSA might be applicable.[31] Specifically, the right to freedom of speech (Art. 11 of the Charter), the right to Freedom of assembly and association (Art. 12 of the Charter), the Right to vote and stand as a candidate at the elections to the European Parliament (Art. 39 of the Charter) and the parallel protection for municipal elections in Art 40.

Regarding advertising, the DSA furthermore defines minimal general rules for advertising online in Art 26 DSA. This provision demands the identifiable and consistent indication that the content is labeled as advertising content (Art. 26 (1) (a) DSA), for whom the advertising is presented (b), who paid for the ad (c), and the main parameters used in the process to identify the audience of the ad (information about aggregated targeting criteria) and the possibility to adjust those according to lit. d. According to Art. 26 (3) in conjunction with Art 4 (4) and Art 9 (1) GDPR[32] shall be prohibited. This may include targeting based on "[. . . ] racial or ethnic origin, political opinions, religious or philosophical beliefs, or trade union membership, and the processing of genetic data, biometric data for the purpose of uniquely identifying a natural person, data concerning health or data concerning a natural person's sex life or sexual orientation". Recital 95 DSA additionally stresses the need for information about targeting criteria and the relationship to vulnerable groups and situations. A demand echoed in the Delegated Regulation for testing systemic risks on these marginalized groups.[33]

---

[30] COLLIER, U.S. celebrities were tricked into recording videos later doctored into anti-Zelenskyy propaganda https://www.nbcnews.com/tech/misinformation/celebrities-tricked-cameo-videos-russia-ukraine-propoganda-elijah-wood-rcna128331 (accessed 11 November 2024), 2023.
[31] Charter of Fundamental Rights of the European Union OJ C 326.
[32] Regulation (EU) 2016/679 of the European Parliament and of the Council of 27 April 2016 on the protection of natural persons with regard to the processing of personal data and on the free movement of such data, and repealing Directive 95/46/EC (General Data Protection Regulation) (Text with EEA relevance) OJ L 119.

Additionally, Art 39 DSA is creating novel rules to increase ad transparency through advertising repositories. Content presented in the repository has to indicate at least the ad itself (Art 39 (2) (a) DSA), for whom the ad is presented (b), who has paid for the ad (c), the duration of presentation (d), information about the parameters of targeting (inclusion and exclusion of groups to the desired audience) of groups (e), commercial communication (f), and the total number of views broken down by Member State (g).

If the advertising content, however, would violate the terms and conditions (Art 14 DSA) of the platform or be regarded as illegal according to Art 3 (h) DSA it should not be included in the repository but should be indicated within the transparency mechanism referred to as Statement of Reason according to Art 17 DSA or within the reporting of received orders from public authorities according to Art 9 DSA. This information covers the reporting on content moderation decisions for platforms at a meta-level. According to the regulation, codes of conduct should complement the compliance measures used in the domain of online advertising and the systemic risks that could be faced by platforms (Art 45 DSA). Such codes should provide key performance indicators as measurements of compliance according to Art 45 para 3 DSA. Systemic risks further create the potential invitation to make new codes of conduct according to Art 45 (2) DSA. Should a platform fail to address the systemic risks adequately, the Commission, the Board, according to Art. 61 DSA, and the relevant signatories (e.g. Who Targets Me for the Code of Practice on Disinformation)[34] will be consulted for repairing the failed attempt (Art 45 (4) DSA).

More specifically, Art 46 DSA defines voluntary and tailored rules for online advertising to increase the need for transparency, highlighting Art 26 and 39 DSA. Additionally, Art 46 (2) (c) DSA includes information on the monetization of data from signatories. This need for standardization is also addressed in Rec. 107 DSA and the relation to the technical implementation of such compliance demands.

## 3.2. Regulation on Transparency and Targeting of Political Advertising

The Regulation on Transparency and Targeting of Political Advertising (TTPA) sets out crucial provisions for safeguarding electoral integrity and transparency in political advertising within the digital domain. The TTPA mandates that political advertisements be clearly identifiable and that platforms provide transparency on ad targeting, financing, and sponsorship. It addresses systemic risks by defining guidelines that ensure voters have access to reliable, unbiased information and can easily identify the sources and funding behind political messaging.[35]

Under the TTPA, all political ads must display the sponsor's name and include sufficient information to identify the individual or entity paying for the advertisement, ensuring transparency in sponsorship (Art 3(10) TTPA, Art 8(1) (b) TTPA). Platforms are required to maintain publicly accessible, searchable repositories of political advertisements, detailing the sponsor, the duration of each ad, the aggregated targeting criteria, and information on who viewed the ad (Art 13 TTPA, Art 39 DSA). This repository allows both the public and auditors to access political ad data, providing oversight and enhancing transparency around targeted political campaigns.

---

[33] Delegated Regulation supplementing Regulation (EU) 2022/2065 of the European Parliament and of the Council, by laying down rules on the performance of audits for very large online platforms and very large online search engines, 20.10.2023.

[34] The Strengthened Code of Practice on Disinformation. https://cmpf.eui.eu/event/strengthened-code-of-practice-on-disinformation/ (accessed on 14 November, 2024), 2022.

[35] EUROPEAN COMMISSION, Commission on Guidelines for providers of Very Large Online Platforms and Very Large Online Search Engines on the mitigation of systemic risks for electoral processes pursuant to the Digital Services Act. https://digital-strategy.ec.europa.eu/en/library/guidelines-providers-vlops-and-vloses-mitigation-systemic-risks-electoral-processes (accessed on 19 November 2024), 2024.

The TTPA further requires VLOPs and VLOSEs to clearly label political ads, distinguishing them from other content. Influencers promoting political content must also declare their sponsorships, with platforms providing functionalities for these declarations. This measure ensures that audiences can differentiate organic content from paid political advertising and understand the affiliations behind the content.

An essential component of the TTPA is transparency in targeting criteria. Platforms must disclose the criteria used for ad targeting, allowing users to understand why certain ads are shown to them, especially where ad targeting includes demographic or interest-based segmentation (Art 3(11) TTPA, Art 8(1) (f) TTPA). Platforms must also uphold ethical targeting practices by providing a "why am I seeing this ad" feature that offers users a straightforward explanation of why they were targeted for specific political messages, further reinforcing transparency in ad delivery.

The TTPA also addresses emerging risks associated with synthetic media, mandating that generative AI content, especially deepfakes, be appropriately labeled. Such labeling helps prevent the spread of misinformation and ensures that voters are not misled by synthetic representations of political figures or events (see also Guideline Sec. 3.3). Media literacy initiatives are also recommended, equipping users with knowledge on how to critically evaluate digital content, particularly AI-generated material. These initiatives support the democratic process by promoting informed decision-making among voters.

Through the TTPA's comprehensive approach, the EU aims to create a transparent, accountable environment for digital political advertising. By defining clear guidelines for sponsorship disclosure, ad targeting transparency, influencer accountability, and synthetic media labeling, the TTPA contributes significantly to maintaining the integrity of digital electoral processes.

### 3.3. The Commission's Guidelines on the Mitigation of Systemic Risks for Electoral Processes Pursuant to the DSA

The European Commission's Guidelines on Mitigation of Systemic Risks for Electoral Processes under the Digital Services Act (G-E–DSA) aim to address digital risks that may compromise the integrity of electoral processes, particularly risks posed by VLOPs and VLOSEs. The guidelines outline best practices and recommended measures that VLOPs and VLOSEs are encouraged to adopt to manage risks related to disinformation, generative AI, and manipulative advertising. These measures are intended to protect public discourse, support transparency, and uphold electoral integrity.

In accordance with Article 35(3) of the DSA, VLOPs and VLOSEs should integrate these guidelines into their risk mitigation frameworks. The guidelines emphasize the importance of transparency in political advertising, requiring that ads be clearly labeled and that platforms maintain accessible ad repositories. These repositories should detail essential information, including the identity of the sponsor, targeting criteria, and the reach of each ad. This transparency enables users to understand the source and intent of the content they encounter, while helping regulators and auditors monitor the digital landscape during election periods.

Generative AI represents a particularly complex risk, and the guidelines encourage platforms to adopt advanced measures for identifying and labeling AI-generated content. Using watermarks, metadata, and other technical tools, platforms can ensure that AI-generated content is authenticated, which is critical in political contexts to prevent the spread of misleading information that could manipulate voter perceptions. To further minimize these risks, platforms are advised to proactively inform users about potential inaccuracies in AI-generated content and to recommend verification through reliable sources. Additionally, the guidelines suggest platforms conduct regular testing, including red-teaming exercises with both internal teams and external experts, to identify and control unintended effects of generative AI content before public release.

Political advertising transparency is a cornerstone of the G-E–DSA, aligning closely with the TTPA. The guidelines mandate that VLOPs and VLOSEs label political ads in a way that is clear, visible, and identifiable in real-time. For influencer-based political content, platforms must provide functionalities that allow influencers to transparently declare sponsorships, ensuring that users can distinguish between organic and paid political messaging.

In addition to labeling, platforms are encouraged to respect 'silence periods', which are times when political advertising is restricted under Member State laws to prevent undue influence on voters immediately before elections. The guidelines also advocate for media literacy initiatives that educate users on recognizing manipulated or AI-generated content, especially when it involves political or election-related information. These initiatives are seen as crucial steps to empower users with the skills needed to critically assess the content they encounter online.

By implementing these guidelines, platforms can more effectively identify and manage systemic risks associated with digital electoral processes. Through measures like transparency in political advertising, monitoring of AI-generated content, enforcement of silence periods, and media literacy initiatives, the G-E–DSA contributes to creating a safer digital environment that respects democratic principles and empowers users in an increasingly complex information landscape.

## 4. Methodology

This study employs a legal doctrinal research approach to develop a comprehensive taxonomy for assessing compliance with the EU's Digital Services Act (DSA), the Transparency and Targeting of Political Advertising Regulation (TTPA) and the Commission's Guidelines on the Mitigation of Systemic Risks for Electoral Processes (G-E–DSA). Building on prior work[36] and aligning with regulatory obligations, we systematically structure a taxonomy that addresses key compliance requirements, annotates risk categories, and establishes content analysis codes specific to electoral integrity and transparency in political advertising.

We aim to contribute to regulatory compliance in the context of the EU by highlighting key provisions in the three legal sources and showing how they complete each other. Furthermore, we want to contribute to the systemic risk assessment and external audit process, according to Art. 34-35 and Art. 37 DSA by proposing a codebook specifying content annotation classes, especially targeting the systemic risks addressed in Art. 34(1)(c) DSA. These content annotation classes are based on relevant literature and legal norms and build on prior work contextualizing the negative effects on the electoral process as proposed in Art. 34 (1) (c) DSA.[37] The codebook defines the following:

- Code names that can be applied to individual pieces of content (e.g., E-1).
- Sources that influenced the creation of a code or legal provisions (e.g., Kapantai[38], or Art. 39 (2)(a) DSA)
- A qualification that determines what the code should entail (e.g.,hoxes).
- A definition providing a better overview of what is meant by the code and how it should be applied and understood.
- Examples to enable a better-aligned application of the codes and increase intercoder reliability.

---

[36] WAGNER ET AL, Mapping interpretations of the law in online content moderation in Germany, Computer Law & Security Review 55, 106054, 2024.
[37] KÜBLER ET AL, The 2021 German Federal Election on Social Media: Analysing Electoral Risks Created by Twitter and Facebook. Proc. of the 56th Hawaii International Conference on System Sciences (HICSS-56), Maui 2023.
[38] KAPANTAI ET AL, A systematic literature review on disinformation: Toward a unified taxonomical framework, New Media & Society 23, 5, 2021, p. 1301.

We propose content analysis as a qualitative methodology for conducting systemic risk assessments under the DSA in line with Art. 34 DSA and testing systemic risk assessment reports as external independent auditors on a sample basis, according to Art. 37 DSA and the Delegated Regulation.[39]

The development of this taxonomy involves a doctrinal analysis of the regulatory texts, synthesizing legislative language, definitions, and thematic requirements across DSA, TTPA, and G-E–DSA. By cross-referencing legal provisions with existing frameworks on disinformation, political advertising, and content moderation, we derive annotation categories that capture the core regulatory obligations. These classes are relevant to future research for computer scientists to automate risk analysis techniques, as well as for legal and social science to have a quality control mechanism to conduct systemic risk assessments. Each category is explicitly tied to legal provisions, with clear definitions, qualifications, and examples to guide the analysis. For instance, in assessing transparency requirements for political ads, our taxonomy incorporates TTPA's mandates for sponsor identification, targeting criteria disclosure, and ad labeling in real-time, reflecting the regulatory emphasis on ensuring public awareness of the origins and targeting mechanisms behind political content.

Content is annotated based on this taxonomy, which includes distinct classifications for transparency, systemic risks, and ad repository completeness. For example, categories are defined according to transparency requirements (e.g., clear sponsorship labeling, real-time targeting disclosure) and systemic risks as outlined in Art. 34 DSA, explicitly focusing on risks to civic discourse, electoral processes, and fundamental rights. Annotations are refined to account for the G–E–DSA's emphasis on managing synthetic and AI-generated content, labeling requirements for synthetic media, and targeted approaches to disinformation in electoral contexts.

This structured taxonomy allows for the application of content analysis techniques on a diverse sample of political content, drawn from prominent VLOPs identified as central to electoral advertising. Using sampling, we propose to select advertisements and user-generated content from VLOPs, representing a broad cross-section of digital content types, including text (from content and comments), images, and audio. The content sample is annotated in line with the taxonomy, with coders assessing compliance with transparency and systemic risk obligations by applying each annotation category. We measure the consistency and accuracy of our coding process through Krippendorff's Alpha,[40] which allows us to refine our taxonomy through pre-tests and ensures reliability in annotation.

Our taxonomy-based approach culminates in a comprehensive report that evaluates compliance across three main regulatory domains, highlighting transparency, systemic risk, and completeness of the political ad repository. The structured taxonomy standardises our content analysis methodology and supports future comparative research by defining clear categories for the complex regulatory landscape of digital political advertising. By mapping content categories directly to legal provisions, our framework bridges the gap between doctrinal interpretations and empirical content analysis, providing a consistent tool to assess systemic risk mitigation practices, compliance with the DSA and TTPA requirements, and potential impacts on electoral integrity.

This methodology demonstrates a novel use of doctrinal research to operationalise regulatory compliance in digital electoral contexts, emphasising the need for empirical methods that align closely with legal taxonomies. Our findings will contribute to the ongoing regulatory discourse, offering platforms, advertisers, and policymakers guidance in managing transparency and systemic risks, ultimately strengthening protections for democratic processes in the digital age.

---

[39] Delegated Regulation supplementing Regulation (EU) 2022/2065 of the European Parliament and of the Council, by laying down rules on the performance of audits for very large online platforms and very large online search engines, 20.10.2023.

[40] ZAPF ET AL, Measuring inter-rater reliability for nominal data – which coefficients and confidence intervals are appropriate?, BMC Medical Research Methodology 16, p. 93.

## 5. Results: Annotating, Reporting, and Transparency Obligations in Political Content and Advertising

In the context of EU regulatory frameworks such as the DSA, TTPA, and G-E–DSA, platforms face increasing obligations to systematically address and report on risks associated with political content, advertising, and misinformation. These frameworks require platforms, especially VLOPs and VLOSEs, to engage in rigorous monitoring and annotation of content to identify, assess, and mitigate systemic risks, particularly those affecting electoral processes. To facilitate these requirements, the proposed taxonomy is a structured methodology for classifying and annotating political content based on risk factors identified in the DSA, TTPA, and G–E–DSA. Categories within this taxonomy include specific annotations for disinformation, procedural electoral threats, misrepresentation in political advertisements, and implications of generative AI use in political contexts. The taxonomy supports compliance by enabling researchers to label and classify content, ensuring transparency in advertising practices, and adherence to reporting requirements for political ads and synthetic media.

Furthermore, the taxonomy provides a foundational structure for reporting obligations. Article 39 of the DSA mandates the creation of ad repositories with detailed information on the nature, targeting, and sponsorship of political advertisements. Our framework defines categories for annotation, facilitates content analysis across various media forms, including text, image, video, and audio, and enables platforms to achieve regulatory compliance while preserving transparency in political discourse.

Through systematic annotation, this framework aligns with EU obligations on transparency, enabling a more accountable, transparent, and equitable political advertising ecosystem. By using the taxonomy, platforms and regulators can more effectively monitor and assess compliance with transparency requirements, ultimately fostering a secure and resilient digital electoral environment.

We developed nine new annotation categories, specifically focusing on the need for these three regulations to be laid out for annotating, especially political advertising content. By selecting these new categories Ro 1-9, we open our coding framework to encounter new regulatory and compliance challenges for testing systemic risks in the context of 'negative consequences for the electoral process.' The coding guideline developed in prior work can be used as an annotation framework to use the adapted taxonomy of this text.[41]

| Category Name | Definition |
| --- | --- |
| Ro-1: Gender-based violence | "Violence against women" means gender-based violence, that is directed against a woman or a girl because she is a woman or a girl or that affects women or girls disproportionately, including all acts of such violence that result in, or are likely to result in, physical, sexual, psychological or economic harm or suffering, including threats of such acts, coercion or arbitrary deprivation of liberty, whether occurring in public or in private life. |
| Ro-2: Sponsorship and identification of sponsors | Ensures all political ads display the sponsor's name and details needed to identify the person who paid for the ad. |

---

[41] WAGNER ET AL, Mapping interpretations of the law in online content moderation in Germany, Computer Law & Security Review 55, 106054, 2024.

| Ro-3: Targeting Criteria Disclosure | Platforms must disclose targeting criteria for political ads, allowing users to understand why they were targeted. |
|---|---|
| Ro-4: Ad Repository Requirements | Platforms must maintain a publicly accessible repository of political ads, including sponsor and targeting details. Platforms must provide transparency around algorithms used in ad targeting, allowing users to understand how content is distributed. |
| Ro-5: Political Ad Labelling | Requires clear, prominent labeling of political ads in real-time to distinguish them as political content. |
| Ro-6: Influencer Political Content | Influencers must declare political content and platforms must label it, ensuring clear disclosure of sponsorships. |
| Ro-7: Labelling of Synthetic Media | Platforms must label synthetic media (e.g., deepfakes, AI-generated misinformation) in political ads to prevent misinformation during elections. |
| Ro-8: Silence Period Enforcement | Platforms must respect national silence periods (ad bans before elections) to avoid undue influence on voters. |
| Ro-9: Monitoring and Fact-checking | Platforms must label content identified by fact-checkers as disinformation, helping users make informed decisions. |

**Table 1: Overview of the newly added categories, considering the negative consequences of the electoral process. For detailed sources, see the Annex.**

We propose using a cluster sampling method in elections or a random sampling method to create a representative dataset of relevant political (advertising) content for annotation.[42] Using content annotation and sampling together, we believe that empirical legal studies on new regulatory challenges can be tackled interdisciplinary while still being flexible enough to incorporate new categories for emerging risks and online harms. By using legal doctrinal methodologies to develop new coding frameworks for specific content domains, we contribute to the ongoing debate on evaluating and measuring compliance within the intersection of several legal domains.

## 6. Conclusion

This paper introduces a comprehensive taxonomy for political content and advertising under the EU's digital electoral framework, encompassing the DSA, TTPA, and G-E–DSA. By developing a structured approach for annotating, reporting, and monitoring political content, this taxonomy provides a foundational tool for VLOPs, VLOSEs, and regulators to assess systemic risks impacting civic discourse and electoral integrity.

The taxonomy proposed here facilitates a nuanced analysis of political content, identifying risk factors and enhancing transparency in advertising practices. It also serves as a basis for improving content moderation strategies, contributing to a more secure and informed digital public sphere. By systematically addressing the regulatory demands, this approach not only aids platforms in meeting their compliance obligations but also strengthens public trust in the integrity of the electoral process within digital environments.

---

[42] WAGNER ET AL, Mapping interpretations of the law in online content moderation in Germany, Computer Law & Security Review 55, 106054, 2024; KÜBLER ET AL, The 2021 German Federal Election on Social Media: Analysing Electoral Risks Created by Twitter and Facebook. Proc. of the 56th Hawaii International Conference on System Sciences (HICSS-56), Maui 2023.

This approach does present some limitations. Within this text the definition of online platforms is used to refer to digital service providers not necessarily in the same understanding as the DSA, since the other regulations analyzed in this text do not share the same differentiation into hosting providers, online platforms, VLOPs or VLOSEs. We will, however, refer to the definitions outlined in the DSA in the correlating sections.

Overall, this work advances the field of digital regulatory compliance by providing a detailed framework for risk assessment, transparency, and digital practices in political content and advertising. Future work will focus on refining these annotation categories in response to emerging risks and on conducting empirical test to validate and operationalize the taxonomy in various electoral contexts.

## 7. Annex: Codebook

### 7.1 Electoral Rights Categories[43]

| Codename & Source | Definition | Example |
|---|---|---|
| C-1: Candidates - Right to stand for elections | Actors interested in restricting electoral competition to ensure electoral victory (of the government) might employ false claims that eligibility conditions are not given in order to exclude certain candidates. | Fabricated criminal allegations. |
| C-2: Candidates - Electoral registration of candidates and parties | Actors interested in restricting electoral competition to ensure electoral victory (of the government) might make erroneous claims that registration requirements are not fulfilled to exclude certain candidates and parties. | Disinformation on changed conditions for the collection of signatures due to the pandemic. |
| C-3: Candidates - Electoral campaign | Actors interested in harming/promoting certain candidates or parties or increasing social and political divisions in society spread malinformation on the private lives of candidates, or disinformation on political intentions, connections and activities of candidates and parties, or false allegations of violating campaign rules in order to defame candidates and parties, manipulate public opinion or influence voting behavior. | Translation by the authors: "Martin Sichert: Billionaire George Soros is known to want to make the world "better" in his favor by imposing an Eco-dictatorship on the people under his control. every year at the world economic forum in Davos..." […] "Bearbock (German politician) undergoes 5-year training program at the world economic forum. There, influential people are brought in to implement a worldwide political and social "change" in the sense of Soros and co. Translation by the authors: |

---

[43] Categories adopted from KÜBLER ET AL, The 2021 German Federal Election on Social Media: Analysing Electoral Risks Created by Twitter and Facebook. Proc. of the 56th Hawaii International Conference on System Sciences (HICSS-56), Maui 2023.

| | | | "Attila Hildmann: Bearbock verurteilte angriffe der hamas auf israel: 'Wir stehen an der Seite der Israelis' […] Bearbock stands on the side of Israel because Bearbock is Jewish" |
|---|---|---|---|
| | E-4: Candidates - Election polls | Actors interested in (de-)legitimizing the elections or harming/promoting certain candidates or parties publish fictitious, false or supportive election polls to (de-) mobilize voters and/or influence both voter turnout and voters' decisions. | Circulation of non-representative or fabricated election polls, or the intentional misinterpretation of representative polls |
| | I-1: Integrity - Voter registration | Actors interested in delegitimizing the elections or actors unwilling to accept electoral results might make undocumented claims that for example "dead voters" are voting to create distrust in the electoral process and to questioning electoral results. | Translation by the authors: "If the federal election takes place by postal vote and many dead people can vote for him, as was the case with Biden, Scholz will certainly become the next chancellor." This disinformation is inspired by similar claims made in the USA during the 2020 election, even though the probability of dead people voting is comparatively low in Germany due to the link between the voter registry and the local resident registry, which does not exist in the USA. |
| | I-2: Integrity - Voting | Actors interested in delegitimizing the election or influencing voting behavior might spread rumors of non-secrecy of voting, sanctions on (non-)voters, electoral violence, as well as last-minute disinformation on election day (e.g. drop-out of candidates), or rumors that voting system is not safe and can be manipulated to create distrust and an atmosphere of fear, influencing voting behaviour. After the election, electoral contenders, particularly defeated candidates, and their supporters can make undocumented claims on systematic vote buying, ballot-box stuffing etc. to | Claims that mail-in votes are inherently fraudulent. |

| | | |
|---|---|---|
| | create distrust, justifying electoral defeat, and questioning electoral results, as well as delegitimizing elections in general and encouraging electoral protests. | |
| I-3: Integrity - Counting and notification | Election losers and their supporters make undocumented claims on lost ballot boxes, and non-counted votes, or the manipulation of vote counts and election protocols etc. to justify electoral defeat, question electoral results and delegitimizing elections, encouraging electoral protests. | Under the hashtag \#Wahlbetrug, it was alleged that election workers had devalued AfD votes, i.e. made them invalid. The former AfD parliamentary group leader of the state parliament of Saxony-Anhalt and now ex-AfD politician André Poggenburg published a corresponding tweet. To underline his claim, he shared the photo of another tweet in which a user pretends to be an election worker in Saxony-Anhalt and attaches a photo on which election workers can be seen counting votes.<br>The photo was not taken during the vote count for the state election in Saxony-Anhalt, but comes from Washoe County, Nevada, taken during the US election in November 2020, as the dts news agency, from which the photo was taken, confirmed to \#Faktenfuchs. This is evident in the community's logo on the white column. The logo is painted over in red in the tweet, as are the faces of the election workers.<br>Translation by the authors: "Almost 18,000 invalid ballot papers!!!???? I've been a poll worker several times. A maximum of 5 invalid votes per district. That stinks to high heaven! \#Voting fraud" |
| I-4: Integrity - Publishing of electoral results | Election losers and their supporters make undocumented claims that late publication or changes of results as reported on election night indicate electoral fraud to justify electoral defeat, delegitimize democratic elections and encourage electoral protests. | On election night, Trump appeared to have a strong "lead" in Pennsylvania, and he expressed confidence he would win the state. He even wrongly "claimed" Pennsylvania's Electoral College votes in a tweet before the race had been decided. But in the days ahead, as more mail ballots were processed and counted, Joe Biden pulled ahead and ultimately won the state by 81,000 votes, or about 1%. The slow counting of mail ballots, and the way it eroded Trump's early advantage, was the direct and expected result of Pennsylvania's election rules — not fraud. The counting process took longer than in other states because elections officials were prohibited from starting their work before Election Day. |
| I-5: Integrity - Electoral results | Election losers and their supporters make undocumented claims on electoral fraud to justify electoral defeat, delegitimize democratic election and encourage electoral protests. | Translation by the authors: "letter election has risen from 14 to 30 percent and the \#CDU has simultaneously risen to almost 37 percent. Coincidences do happen... Anyone who believes that still believes in Father Christmas. The establishment will stop at nothing, not even electoral fraud. \#electionfraud \#Itwsa21" |
| I-6: Integrity - Electoral observation | Government and its national/international supporters; government-friendly media promote false claims by pro-government election observers to legitimize non-democratic elections. Partisan election observers and media; foreign governments make false claims by partisan election observers (e.g. exaggerating the importance of minor | Unsubstantiated claims of a lack of independent oversight of the voting process and the vote count. |

| | irregularities) to delegitimizing democratic elections. | |

## 7.2 Regulation of political content

| Codename & Source | Definition | Example |
|---|---|---|
| Ro-1 Gender-based violence (Rec 81 DSA, (Art 34(1)(d) DSA), Art. 3-9 Directive combating violence against women and domestic violence[44]) | Rec 81 DSA, (Art 34(1)(d) DSA), Art. 3 Directive combating violence against women and domestic violence Female genital mutilation, Art. 4 Forced marriage, Art. 5 Non-consensual sharing of intimate or manipulated material, Art 6 Cyber stalking, Art 7 Cyber harassment, Art 8 Cyber incitement to violence or hatred, Art 9 Inciting, aiding and abetting and attempt<br><br>Rec. 83 "A fourth category of risks stems from similar concerns relating to the design, functioning or use, including through manipulation, of very large online platforms and of very large online search engines with an actual or foreseeable negative effect on the protection of public health, minors and serious negative consequences to a person's physical and mental well-being, or on gender-based violence. Such risks may also stem from coordinated disinformation campaigns related to public health, or from online interface design that may stimulate behavioural addictions of recipients of the service." | For example the case when the intentional act of distributing, publishing, or otherwise making accessible intimate images, videos, or audio recordings of an individual without their explicit consent is performed. This includes material originally obtained with consent but shared beyond its agreed purpose, as well as content that has been digitally altered or manipulated to degrade, humiliate, or harm the individual, e.g. like in revenge pron. |

---

[44] Directive (EU) 2024/1385 of the European Parliament and of the Council of 14 May 2024 on combating violence against women and domestic violence.

| | Art 2 (a) and (b), Directive (2024/1385): "(a) "'violence against women' means all acts of gender-based violence directed against a woman or a girl because she is a woman or a girl or that affect women or girls disproportionately, that result in or are likely to result in physical, sexual, psychological or economic harm or suffering, including threats of such acts, coercion or arbitrary deprivation of liberty, whether occurring in public or in private life; […] 'domestic violence' means all acts of physical, sexual, psychological or economic violence that occur within the family or domestic unit, irrespective of biological or legal family ties, or between former or current spouses or partners, whether or not the offender shares or has shared a residence with the victim;" | |
|---|---|---|
| Ro-2 Sponsorship and identification of sponsors (Art 3 (10) TTPA, Art 8 (1) (b) TTPA, Art 9 (1) (e) TTPA, Art 10 TTPA, Art 11 (1) (b), Art 12 (1) (a-b) TTPA ) | Ensures all political ads display the sponsor's name and details needed to identify the person who paid for the ad. | A political ad without identifiable information about the sponsor. |
| Ro-3 Targeting Criteria Disclosure (Art 3 (11) TTPA, Art 8 (1) (f) TTPA, Art 10 TTPA, Art 11 (1) (d) TTPA, Art 12 (1) (c) TTPA, Art 18 TTPA, Art 19 TTPA, Art 35 (3) DSA) | Platforms must disclose targeting criteria for political ads, allowing users to understand why they were targeted. | An ad targeting specific demographics without disclosing the criteria, e.g. women in a certain zip code area. |
| Ro-4: Ad Repository Requirements (Art 13 TTPA, Art 39 DSA) | Platforms must maintain a publicly accessible repository of political ads, including sponsor and targeting details. Platforms must provide transparency around algorithms used in ad | A platform that doesn't archive political ads for public access. |

| | targeting, allowing users to understand how content is distributed. | |
|---|---|---|
| Ro-5: Political Ad Labelling(Art 26 DSA, Art 39 DSA, Art 11 TTPA, Art 28 TTPA) | Requires clear, prominent labeling of political ads in real-time to distinguish them as political content. | Political ads lacking labels, making it unclear they're paid content. |
| Ro-6: Influencer Political Content (G-E–DSA Sec. 3.2.1(f)) | Influencers must declare political content and platforms must label it, ensuring clear disclosure of sponsorships. | Influencer posts endorsing candidates without sponsorship disclosure. |
| Ro-7: Labelling of Synthetic Media (DSA Art. 39, G-E–DSA Sec. 3.2.3) | Platforms must label synthetic media (e.g., deepfakes, AI-generated misinformation) in political ads to prevent misinformation during elections. | AI-generated videos of candidates without clear labeling as synthetic. |
| Ro-8: Silence Period Enforcement (Rec. 14) | Platforms must respect national silence periods (ad bans before elections) to avoid undue influence on voters. | A political ad shown during the silence period prior to voting day. |
| Ro-9: Monitoring and Fact-checking (G-E–DSA Sec. 3.3) | Platforms must label content identified by fact-checkers as disinformation, helping users make informed decisions. | Content spreading misinformation without a fact-check label. |

## 7.3 Disinformation Content[45]

| Codename & Source | Definition | Example |
|---|---|---|
| D-1 Fabricated | Stories that completely lack any factual base, are 100% false. The intention is to deceive and cause harm. One of the most severe types of fabrication adopts the style of news articles so the recipients believe it is legitimate. Could be text but also in visual format. | Examples include the news articles that are completely invented, often used to delegitimize a political candidate during an election. |
| D-2 Imposter | Genuine sources that are impersonated with false, made-up sources to support a false narrative. It is actually very misleading since source or author is considered a | Examples include Facebook sites set up with very similar looking logos and names of very known media, such as the Guardian or the BILD. |

---

[45] Categories and definitions are adopted from KAPANTAI ET AL, A systematic literature review on disinformation: Toward a unified taxonomical framework, New Media & Society 23, 5, 2021, p. 1325.

| | | |
|---|---|---|
| | great criterion of verifying credibility (use of journalists name/ logo /branding of mimic URLs). | |
| D-3 Conspiracy theories | Stories without factual base as there is no established baseline for truth. They usually explain important events as secret plots by the government or powerful individuals. Conspiracies are, by definition, difficult to verify as true or false, and they are typically originated by people who believe them to be true. Evidence that refutes the conspiracy is regarded as further proof of the conspiracy. Some conspiracy theories may have damaging ripple-effects. | Examples for conspiracies include the alleged planning and orchestrating of the 9/11 attacks by the US government, or the belief that politicians and other powerful people are reptilians and/or aliens. |
| D-4 Hoaxes | Relatively complex and large-scale fabrications which may include deceptions that go beyond the scope of fun or scam and cause material loss or harm to the victim. They contain facts that are either false or inaccurate and are presented as legitimate facts. This category is also known in the research community either as half-truth or factoid stories able to convince readers of the validity of a paranoia-fueled story. | Famous hoaxes often include the distribution of a rumour that a celebrity or person of public interest has died. |
| D-6 Rumours | Refers to stories whose truthfulness is ambiguous or never confirmed (gossip, innuendo, unverified claims). This kind of false information is widely propagated on online social networks. | Rumours can include a wide variety of information, for instance, that Facebook will start charging their users or that a Nigerian Astronaut is lost in space and needs money to return to earth. |
| D-7 Clickbait | Sources that provide generally credible or dubious factual content but deliberately use exaggerated, misleading, and unverified headlines and thumbnails to lure readers to open the intended Web page. The goal is to increase their traffic for profit, popularity, or | These types of articles often include mundane content such as "do not eat this if you want to loose weight" or "the world was shocked to learn about the truth of…". |
| | sensationalization. Once the reader is there, the content rarely satisfies their interest. | |
| D-8  Misleading connection | Misleading use of information to frame an issue or individual. When headlines, visuals, or captions do not support the content. Separate parts of source information may be factual but are presented using wrong connection (context/content). | These are often related to decontextualisation. A famous example was a photo by prince William talking to photographers, the photo and text made it seem as if he was using an offensive hand gesture, however, from a different angle, the situation was proven to be completely innocent. |
| D-9  Trolling | The act of deliberately posting offensive or inflammatory content to an online community with the intent of provoking readers or disrupting conversation. Today, the term "troll" is most often used to refer to any person harassing or insulting others online. | Trolling can often be done by a group of people using distinctive symbolism to identify as trolls (such as Pepé the Frog). |
| D-10  Pseudoscience | Information that misrepresents real scientific studies with dubious or false claims. Often contradicts experts. Promotes metaphysics, naturalistic fallacies, and others. The actors hijack scientific legitimacy for profit or fame. | Famous examples are the claims that vaccines cause autism or "proofs" that the earth is actually flat. |